\documentclass{config/ifacconf}
\usepackage{graphicx}      
\usepackage{natbib}        
 \usepackage{amsmath}
 \usepackage{xcolor}
 \usepackage[version=4]{mhchem} 
 \usepackage[table]{xcolor}

\begin{document}
\begin{frontmatter}

\title{A simulation- and model-based approach to PI control pairing and tuning for the pyro process in a cement plant
}



\thanks[footnoteinfo]{Corresponding author: J.B. Jørgensen (e-mail: jbjo@dtu.dk)}

\author[2C]{Jan Lorenz Svensen} 
\author[2C]{Steen Hørsholt} 
\author[RAMCO]{Guruprasath Muralidharan}
\author[2C,DTU]{John Bagterp J\o rgensen}

\address[2C]{2-control ApS, DK-7400 Herning, Denmark}
\address[RAMCO]{Ramco Industrial Technology Services, IN-600004 Chennai, India}
\address[DTU]{Technical University of Denmark, DK-2800 Kgs Lyngby, Denmark}

\begin{abstract}
The operation of the pyro process in cement production significantly affects the energy efficiency and sustainability of the cement plant, especially for reductions in carbon dioxide emissions. Hence, pyro process control is essential to obtain efficient and sustainable operation of cement plants. In this paper, we demonstrate how simulations and models can be utilized to evaluate and design control strategies for the pyro section in cement plants. We apply a novel differential algebraic equation (DAE) model for dynamic simulation of the pyro-section in cement plants to design decentralized PI controllers for the pyro-section. We utilize the pyro-process model to evaluate the control structure design. Through linearization of the pyro-process model, we apply the Relative Gain Array (RGA) method to choose and evaluate the pairings of the manipulated variables (MVs) and the controlled variables (CVs). Using simulations of the pyro-section, we generate step responses to estimate transfer models and apply Internal Model Control (IMC) for the tuning of the individual decentralized single-input single-output (SISO) PI controllers. Closed-loop simulations of the PI controllers demonstrate that PI controllers with IMC parameters provide smoother and faster responses compared with manually tuned PI parameters.
\end{abstract}

\end{frontmatter}

\section{Introduction}
Cement plants constitute a vital part of the construction industry and infrastructure of modern civilization. However, they are responsible for 8\% of global \ce{CO2} emissions~\citep{CO2Techreport}.
Thus, improving the balance between economic progress and environmental responsibility is important when using cement for highways, bridges, and cities. To move towards zero \ce{CO2} emission cement plants, a first step is to apply advanced process control (APC) for more efficient operation of existing cement plants, while further steps may be to modify the process and reduce the clinker dependence or capture and store the \ce{CO2} produced in the cement manufacturing process. 
%
Traditional cement is primarily made of clinker. The production of clinker involves mining, blending, grinding, and pyro processing the raw materials. The pyro-process is the main source of emissions and energy consumption. 
Research has studied several alternative strategies \citep{Ramasamy:etal:2023:JPC}, including changing parts of the overall process, such as using electrification and carbon capture \citep{Varnier:etal:2025}.

 A range of advanced control strategies for cement plants have been implemented and documented. These include fuzzy logic control \citep{Holmblad:Ostergaard:1993fuzzylogic}, rule-based neural network control \citep{Bo:etal:1997}, and model predictive control (MPC).
 MPC applications include \cite{Stadler:etal:2011}, \cite{Zanoli:etal:2025AUCC,Zanoli:etal:2015,Zanoli:etal:2023}, \cite{Teja:etal:2016} for kiln and cooler control,  \cite{ZHANG20211319,Zhang:etal:2022:Blend} for raw meal blending and grinding, \cite{Guruprasath:Jorgensen:2009adchem} and \cite{Guruprasath:etal:2010cementMill,Guruprasath:etal:2013} for soft constrained based robust MPC of cement mills, and \cite{Huusom:etal:2005cementmill} for predictive control motivated physically based models of cement grinding units. Several models have been proposed to describe the processes. \cite{Spang:1972} described a reaction-based dynamic kiln model. \cite{Westerlund:1981} provided insight into data-based kiln models using ARMAX models. \cite{Noshirvani:etal:2009}
suggested both Box-Jenkins models and neural-network models.
\cite{Stadler:etal:2011} proposed a mass-based dynamic kiln model. 
\cite{Melitos:etal:2025escape} suggested a steady-state pyro-process model.
\cite{Svensen2024Calciner,Svensen2024Kiln,Svensen2024Cyclone,Svensen2024Cooler,svensen2025FullSim} provided dynamic models for each process section and a full pyro process model.

\begin{figure}
    \centering
    \includegraphics[width=0.94\linewidth]{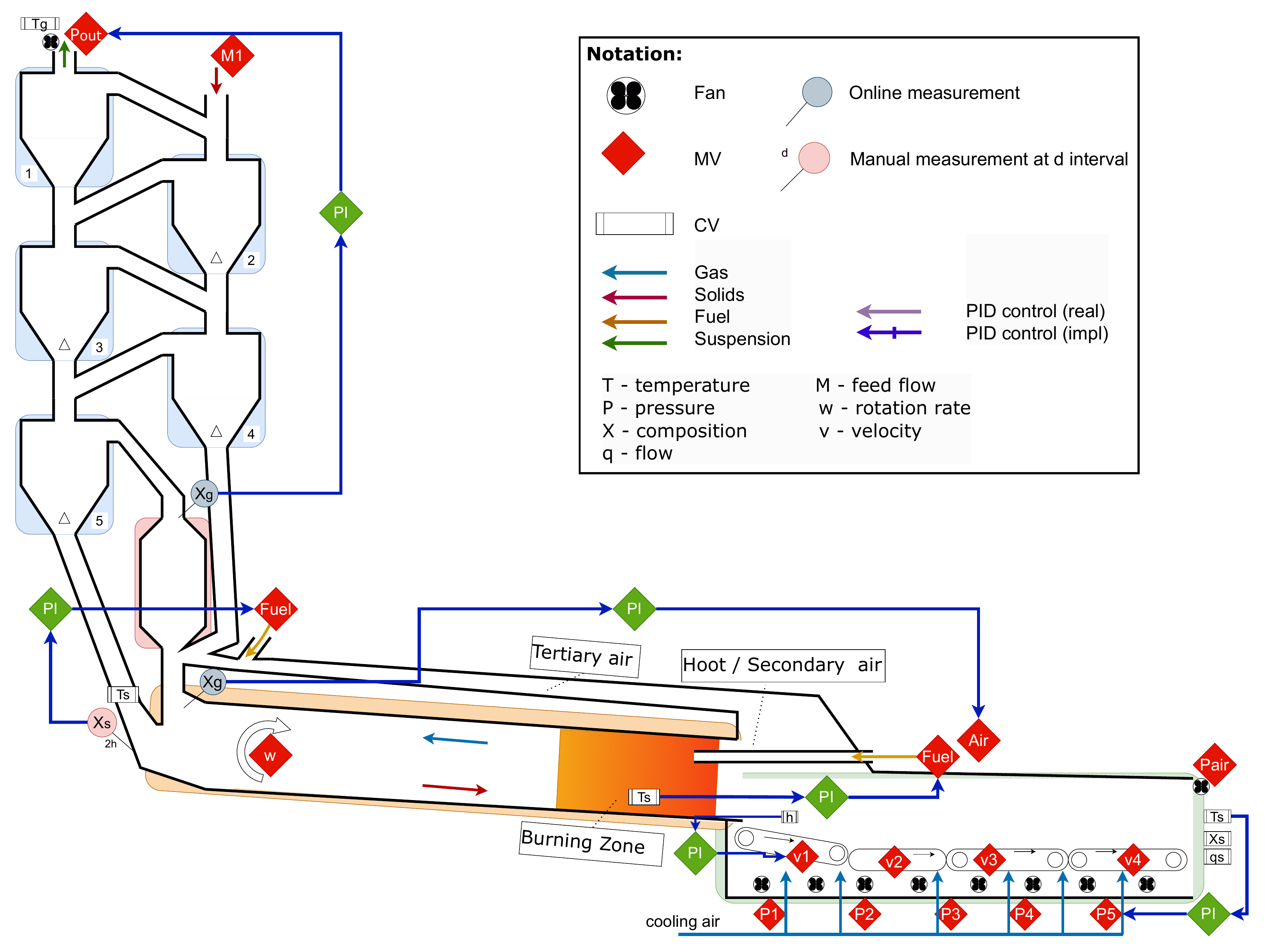}
    \caption{a 5-stage pyro process with PI-control loops for cement production.}
    \label{fig:PIloop}
\end{figure}
This work focuses on the pyro process shown in Figure~\ref{fig:PIloop}. The pyro process consists of a preheating tower with five cyclones and calciner, a kiln, and a cooler. The start-up, shut-down, and operation of the pyro process is complicated and expensive. Any lost production or off-spec product is associated with a large economic penalty. Therefore, testing new control strategies and performing identification experiments (say step response tests) are often not allowed without detailed justification of the risks and benefits. A high-fidelity simulator can provide such justification. We demonstrate the simulator for model-based PI-controller tuning and control structure selection. Such model-based PI-pairing and tuning for the pyro-process in a cement plant is novel.



This paper is organized as follows. Section~\ref{sec:sys} presents the pyro process, Section~\ref{sec:step} describes the IMC control design, Section~\ref{sec:RGA} presents the RGA analysis, and Section~\ref{sec:Conclusion} provides the conclusions.

\section{system and simulator}\label{sec:sys}
Figure~\ref{fig:PIloop} illustrates the layout of the pyro process in cement clinker production. It is a 5-stage single-string plant, meaning that the preheating tower contains 5 separation cyclones in series. 

The raw meal enters near the top of the preheater. On its course through the preheater and calciner, it is heated and calcination occurs. In the kiln, the meal is heated to around 1300-1500$^\circ$C at which it reacts chemically to form the different compounds of cement clinker. Lastly, the clinker passes through the cooler, where it is cooled rapidly to around 100$^\circ$C to stabilize the products.

In the cooler, cooling air is partially circulated back into the process. The first part of the recirculation air flow, the secondary air, is sent to the kiln. The next part, the tertiary air, is sent to the calciner, while the rest escapes to the environment. In the kiln, the secondary air is mixed with the fuel-primary air mixture. Thus, fuel combustion and this recirculated heat provide the heat for the kiln reactions. In the calciner, the kiln gas is mixed with the tertiary air and additional fuel. This heats the feed and facilitates the main part of calcination. The gas flows up through the preheating tower and transfers its heat to the solid feed that flows down.

We use the simulation model described by \cite{svensen2025FullSim}. 
The mathematical model is a differential-algebraic system consisting of 755 differential equations and 157 algebraic equations in the form:
\begin{subequations}
\label{eq:DAE}
\begin{align}
    \dot{x} &= f(x,y,u,d,\theta_f),\\
          0 &= g(x,y,u,d,\theta_g), \\
          z &= h(x,y,u,d,\theta_h).
\end{align}
\end{subequations}
$x$ are the differential states, $y$ is the algebraic states, $u$ is the manipulated variables (MVs), $d$ is the disturbances, $z$ are the controlled variables (CVs), and $\theta$ are the model parameters. 
The DAE model includes thermo-physcial properties, mass and energy transport, reaction stoichiometry and kinetics, and mass and energy balances with appropriate thermodynamic closures.

The MVs are 
\begin{equation}
    u=\begin{bmatrix}
        P_{ph} & F_{f,Ca} &F_{f,K}&F_{1st}& F_{cool}&v_{grate}
    \end{bmatrix}^T.
\end{equation}
$P_{ph}$ is the fan-induced pressure above the preheating tower, $F_{f,Ca}$ is the fuel mass flow going into the calciner, $F_{f,K}$ is the fuel mass flow going into the kiln, $F_{1st}$ is the primary air mass flow send into the kiln with the kiln fuel, $F_{cool}$ are cooling air flow injected below the grate belt of the cooler, and $v_{grate}$ are the grate belt velocities.
For simplicity, the $F_{cool}$ fluxes and $v_{grate}$ velocities are assumed identical across the cooler length.

The CVs are
\begin{equation}
    z=\begin{bmatrix}
        X_{O2,Ca} & X_{CaO} &X_{O2,K}&T_{s,burn}&T_{clinker}&h_{bed}
    \end{bmatrix}^T.
\end{equation}
$X_{O2,Ca}$ and $X_{O2,K}$ are the mole fraction of \ce{O2} at the calciner outlet and kiln inlet, respectively. $X_{CaO}$ is the calcination degree of the feed at the kiln inlet. $T_{s,burn}$ is the clinker temperature in the burning zone of the kiln.
$T_{clinker}$ is the clinker temperature exiting the cooler, and $h_{bed}$ is the height of the clinker bed in the cooler.

Using the Jacobians of the DAE model \eqref{eq:DAE} at steady-state, a standard linear state space model, $(A,B,C,D)$, for the DAE model \eqref{eq:DAE} is obtained by
\begin{subequations}
\begin{align}
    A &= \partial_x f - \partial_y f \left[\partial_y g\right]^{-1} \partial_x g, 
    \\
    B &= \partial_u f - \partial_y f \left[\partial_y g\right]^{-1} \partial_u g, 
    \\
    C &= \partial_x h - \partial_y h \left[\partial_y g\right]^{-1} \partial_x g, 
    \\
    D &= \partial_u h - \partial_y h \left[\partial_y g\right]^{-1} \partial_u g. 
\end{align}
\end{subequations}
The corresponding transfer function, $G(s)$, is computed by
\begin{equation}
    G(s) = C(sI-A)^{-1}B+D.    
\end{equation}
\section{Control and Step responses}\label{sec:step}

We design decentralized PI controllers using a manual trial-and-error tuning method for each PI-parameter as well as by using the simplified IMC (SIMC) tuning method \citep{SKOGESTAD2003291}.
Both type of PI-controllers are simulated as continuous-time controllers that are implemented with limits but without anti-windup.

The manual tuning approach is included to illustrate the usage of the simulator as a digital twin for controller tuning and evaluation. The SIMC tuning approach
highlights the benefits of using the simulator to provide data for model based control design. 

\begin{figure*}
    \centering
    \includegraphics[width=1\linewidth]{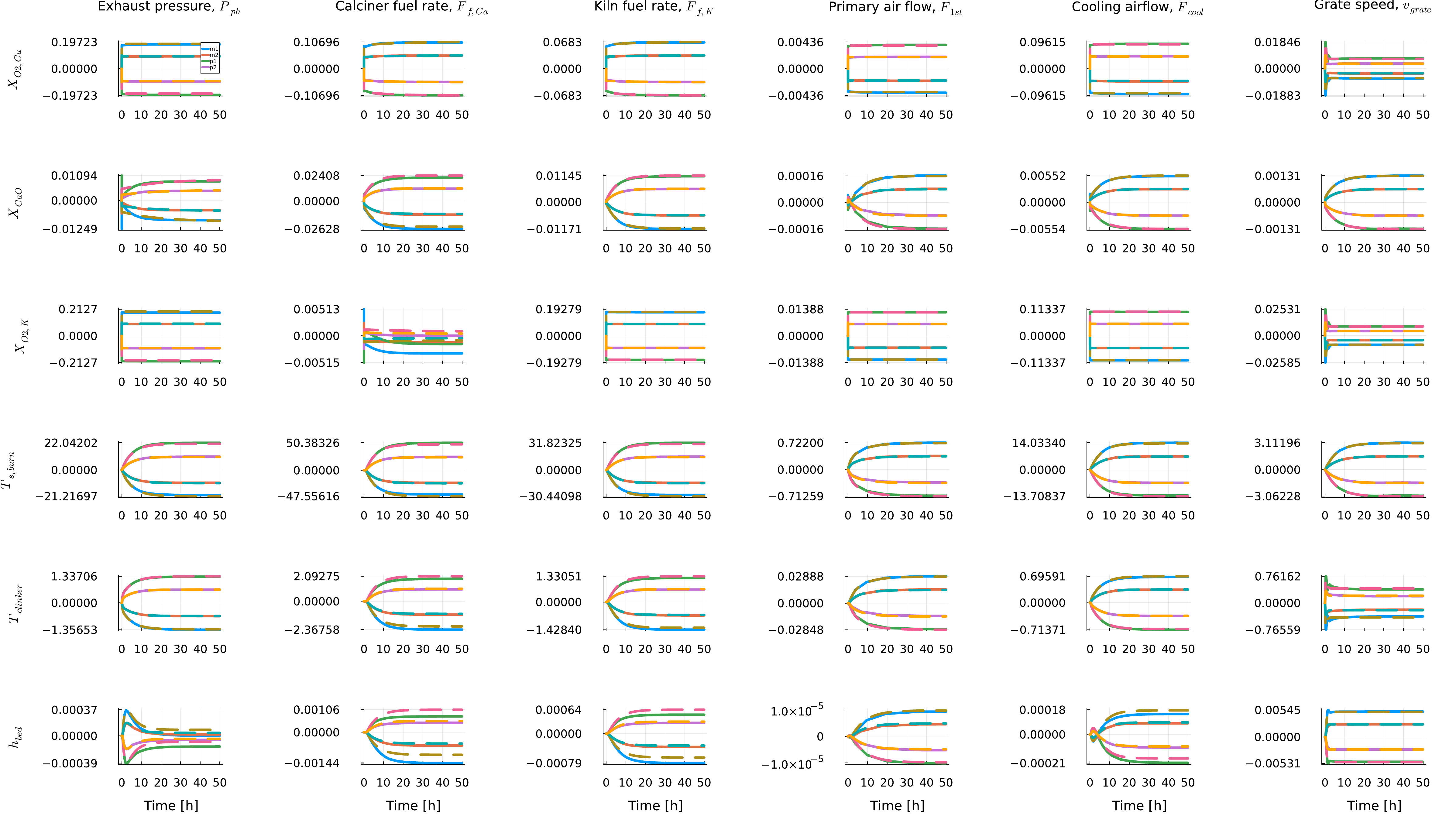}
    \caption{MV-CV step responses: simulated step responses of the simulator for 4 different step sizes (m1: -1\%(blue), m2: -0.5\%(red), p1: 1\%(green), p2: 0.5\%(purple)) and step responses of the estimated model (dashed lines). Each row shows a CV's centered response to the MVs of each column.
    }
    \label{fig:s_2}
\end{figure*}
\begin{figure}
    \centering
    \includegraphics[width=\linewidth]{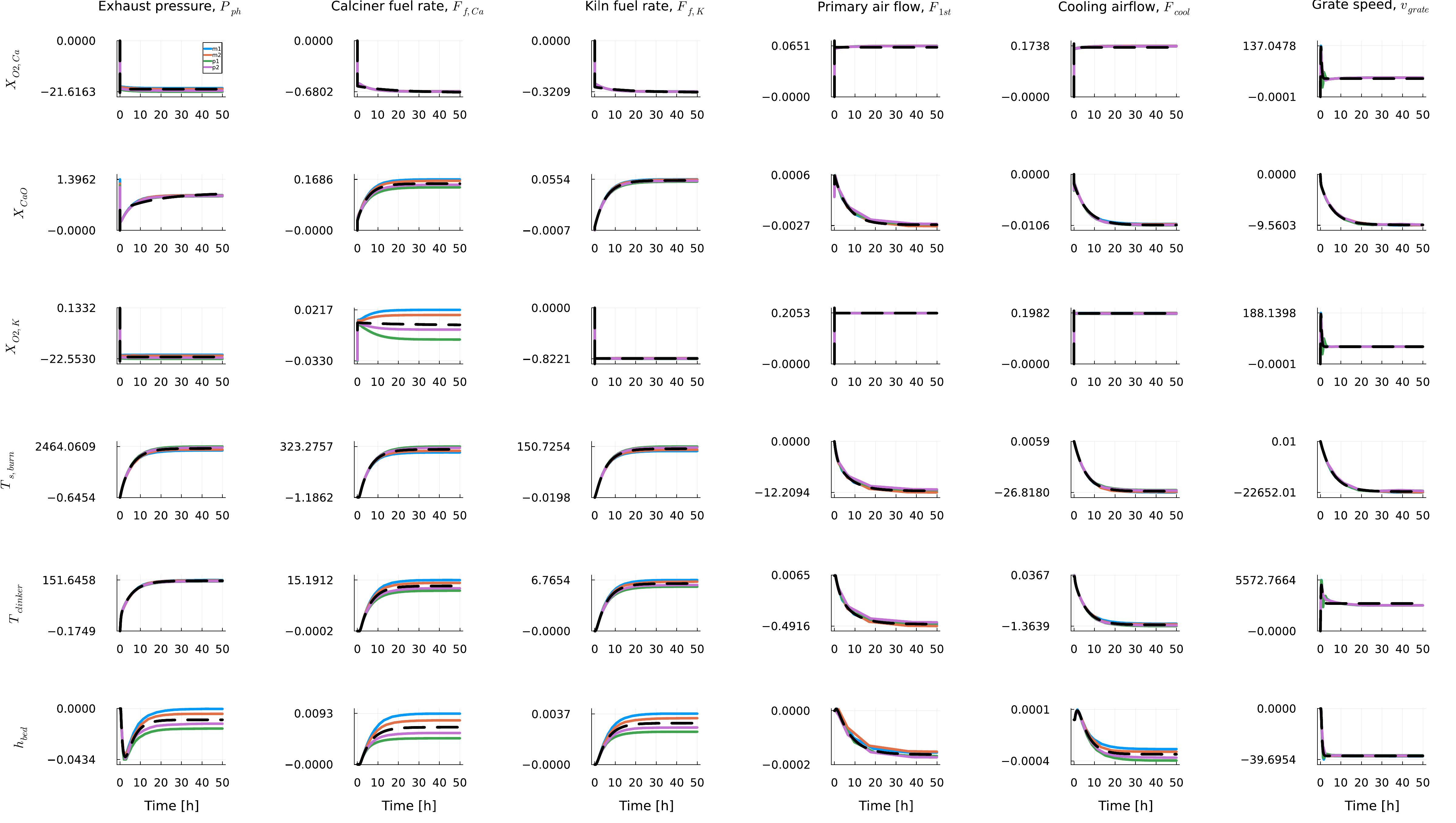}
    \caption{MV-CV normalized step responses for 4 different step sizes (m1: -1\%(blue), m2: -0.5\%(red), p1: 1\%(green), p2: 0.5\%(purple)) and estimated model (dashed lines).}
    \label{fig:norm}
\end{figure}

The SIMC tuning approach needs a transfer function model for each (MV-CV) pair in the control structure. We simulate 4 sets of step responses to generate the data to estimate the models.
The steps were sampled 100 times an hour (i.e. the sampling time is 36 seconds) with steps of $\pm$ 1\% and $\pm$ 0.5\% of steady-state control for each input. 
Figure~\ref{fig:s_2} shows the CV step responses obtained for steps in the MVs. The manual chosen MV-CV control pairs are in the diagonal. Figure~\ref{fig:norm} shows the normalized responses.

\subsection{Transfer function estimation}
Given the shape of the responses in Figure~\ref{fig:s_2} and Figure~\ref{fig:norm}, the data is fitted to a second-order transfer function,
    \begin{align}
        Z_i(s) &= \hat{G}_{ij}(s)U_j(s), \quad \forall i \times j \in CV \times MV,\\
       \hat{G}_{ij}(s) &= K_{ij,0}\frac{\tau^{ij}_zs+1}{(\tau^{ij}_1s+1)(\tau^{ij}_2s+1)}e^{-\tau_{d}^{ij}s}.
    \end{align}
Figure~\ref{fig:norm} shows that for all but a few, the normalized responses show similar dynamic behaviors across the 4 step sizes. The stationary values settle in the same vicinity, with only one pair having shifting signs, ($F_{f,Ca}$,$X_{\ce{O2},K}$).
Given the results, we deem linear models to be a good representation in this range as the nonlinear effects appear to be minimal.

For the estimation of the transfer functions, we normalize the data between the stationary periods $t_0$ and $t_f$,
\begin{align}
        s_{ij}(t) &= \frac{z_{i}(t)-z_{i}(t_0)}{u_{j}(t)-u_{j}(t_0)},
    \end{align}
such that $s_{ij,t}$ corresponds to $\hat{G}_{ij}(s)$. As $U(s) = 1/s$ is the unit step, the normalized step transfer function is
\begin{align}
    \bar{S}_{ij}(s) &= \hat{G}_{ij}(s)\frac{1}{s}.\label{eq:dG(s)}
\end{align}
We fit the data to the time-domain equivalents of \eqref{eq:dG(s)}. The equivalent function depends on the values of $\tau_1$ and $\tau_2$. Generally, the function is given by
\begin{equation}
        \Bar{s}_{ij}(t) = K_h\bigg(1  - \frac{\tau_z^{ij}-\tau_1^{ij}}{\tau_2^{ij}-\tau_1^{ij}}e^{-\frac{\Delta t}{\tau_1^{ij}}} - \frac{\tau_2^{ij}-\tau_z^{ij}}{\tau_2^{ij}-\tau_1^{ij}}e^{-\frac{\Delta t}{\tau_2^{ij}}}\bigg),\label{eq:Gc1}
\end{equation}
but in the case of $\tau_1 = \tau_2$, it is given as
  \begin{align}
        \Bar{s}_{ij}(t) &= K_h\bigg(1 - \bigg(1 + \Delta t\frac{\tau_1^{ij}-\tau_z^{ij}}{(\tau_1^{ij})^2}\bigg)e^{-\frac{\Delta t}{\tau_1^{ij}}}\bigg). \label{eq:Gc2}
    \end{align}
$\Delta t = t-\tau_{d}^{ij}$ is the time offset and $K_h = K_{ij,0}H_d$ is a delayed gain, where $H_d = H(\tau_{ij,d})$ is the heavy-side function.
Figure~\ref{fig:s_2} and Figure~\ref{fig:norm} show the step responses of \eqref{eq:DAE} and of the estimated transfer functions (dashed lines), which clearly fit the data. 
Initial guesses for the stationary gain and delay can be determined by
    \begin{align}
        K_{ij,0} = s_{ij}(t_f), 
    \end{align}
and by evaluating the earliest change:
    \begin{align}
        \tau_{d}^{ij} : |s_{ij}(\tau_{d}^{ij})| > \epsilon, \qquad \epsilon = 10^{-10}.
    \end{align}

The (MV-CV) pairs in the decentralized PI-control structure are ($P_{ph}$-$X_{O2,Ca}$), ($F_{f,Ca}$-$X_{CaO}$), ($F_{f,K}$-$T_{s,burn}$), ($F_{1st}$-$X_{O2,K}$), ($F_{cool}$-$T_{clinker}$), and ($v_{grate}$-$h_{bed}$).
 Exhaust pressure and primary air control local \ce{O2} levels through the air flows. Fuel rates regulate local process conditions, i.e. the calcination degree and the clinker burner temperature. Cooling air regulates the clinker cooling process i.e. the clinker temperature. The grate belt velocity regulates bed height to ensure flow uniformity.

Table~\ref{tab:tfest} shows the estimated parameters for the transfer functions corresponding to the (MV-CV) pairs in the decentralized control structure.
\begin{table}[tb]
    \centering
    \caption{Estimated transfer functions for the (MV-CV) pairs in the decentralized control structure. $\tau$ is in seconds.}
    \scriptsize
    \begin{tabular}{|cc|c|c|c|c|c|}\hline
     MV & CV   & $K_0$ & $\tau_1$ & $\tau_2$ & $\tau_z$ & $\tau_d$\\ \hline
     $P_{ph}$ & $X_{O2,Ca}$ & -20.54   &    7.93   &   7.93  &  21.74  &     0.18\\
    $F_{f,Ca}$ &$X_{CaO}$ &  0.29  &   9.23 & 18103.3     & 3663.1      &    0.04\\
    $F_{f,K}$ & $T_{s,burn}$ &103.35   &  2153.3  &    19191.9   &   1187.8   &    162.02\\
    $F_{1st}$ & $X_{O2,K}$ &0.26   &  2.64  &   16.4    & 32.07      &  0.024\\
    $F_{cool}$& $T_{clinker}$& -4.75 &  1001.1  &   20072.9   &   2694.2   &   1073.9\\
    $v_{grate}$ & $h_{bed}$ & -36.88  &   969.2   &   969.2  &     -0.38 &  920.3\\ \hline
    \end{tabular}
    \label{tab:tfest}
\end{table}

    
\subsection{IMC}
The design of the PI-controller using IMC design \citep{SKOGESTAD2003291}, relies on approximating the general transfer function,
    \begin{equation}
        \begin{split}
            G_0(s) &= k_0\frac{\prod_i^{n_z}(\tau_{z,i}s+1)}{\prod_i^{n_p}(\tau_{p,i}s+1)} e^{-\tau_{d}s},\\  \tau_{k,i}&\geq\tau_{k,j}  \quad \text{for }i\geq j\quad \text{and }k\in \{z,p\},
        \end{split}
    \end{equation}
into a first-order system,
\begin{align}
    G_1(s) = \frac{k}{\tau_{1}s+1} e^{-\tau_d s}.
\end{align}
The IMC rules for the approximation depend on the sign of the zeros, $\tau_z$. For positive $\tau_z$, the rules approximate a zero-pole fraction,
\begin{equation}
        \frac{\tau_zs+1}{\tau_ps+1},
    \end{equation}
as a gain scaling and/or replacement pole. When each positive $\tau_z$ is approximated, the effective delay $\tau_d$ and the first-order poles are computed from the remaining poles, zeros, and delay.
    \begin{align}
        \tau_1 &= \tau_{p,1} + \frac{\tau_{p,2}}{2},\\
        \tau_d &:= \tau_{d} + \frac{\tau_{p,2}}{2} + \sum_{j=3}^{n_p}\tau_{p,j}- \sum_{i\in \tau_{z,i}\leq0}^{n_z}\tau_{z,i}. 
    \end{align}
The PI gains are then computed by
\begin{align}
    K_P = \frac{1}{k}\frac{\tau_1}{\tau_c +\tau_d}, \qquad K_I = \frac{K_P}{\min(\tau_1,4(\tau_c+\tau_d))},
\end{align}
with $-\tau_d<\tau_c<\infty$. In the tuning, we use the standard recommendation of $\tau_c = \tau_d$ as our initial parameter. 
We kept this tuning in the controllers ($P_{ph}$-$X_{O2,Ca}$), ($F_{f,K}$-$T_{s,burn}$), and ($F_{cool}$-$T_{clinker}$), while $\tau_c$ was further adjusted in the remaining control loops.
Table~\ref{tab:PI}, shows the PI gains ($K_P$ and $K_I$) and $\tau_c$ for each pairing.

\begin{table}[tb]
    \centering
    \caption{Pairings and parameters for the decentralized PI-controllers with manual tuning and IMC-based tuning. $\tau_c$ is in seconds.}
    \scriptsize
    \begin{tabular}{|l@{\hspace{-0.05pt}}l|c|c|c|c|c|}\hline
    MV& CV& \multicolumn{2}{|c|}{$K_P$} & \multicolumn{2}{|c|}{$K_I$}& $\tau_c$\\ \hline
          &       & IMC & Manual & IMC & Manual&\\ \hline
    $P_{ph}$ & $\text{ }X_{O2,Ca}$ & -0.0089&-0.01&-0.006&-0.001 & 0.18\\
    $F_{f,Ca}$ &$\text{ }X_{CaO}$ &12.51&0.005&1.35&0.011 & 10.0\\
    $F_{f,K}$ & $\text{ }T_{s,burn}$ & 0.47&0.07&3.6e-4&1.4e-5 & 162.02\\
    $F_{1st}$ & $\text{ }X_{O2,K}$ &0.014&0.008& 0.0069&0.16 & 0.5\\
    $F_{cool}$& $\text{ }T_{clinker}$&-0.32&-2.8e-4&-3.2e-4&-2.8e-4 & 1073.9\\
    $v_{grate}$ & $\text{ }h_{bed}$ &-0.097&-0.02&-6.7e-5&-0.0001 & -1000.0\\ \hline
    \end{tabular}
    \label{tab:PI}
 \end{table}
                   
\begin{figure}
    \centering
    \includegraphics[width=1\linewidth,trim={0cm 2cm 0cm 0.0cm},clip]{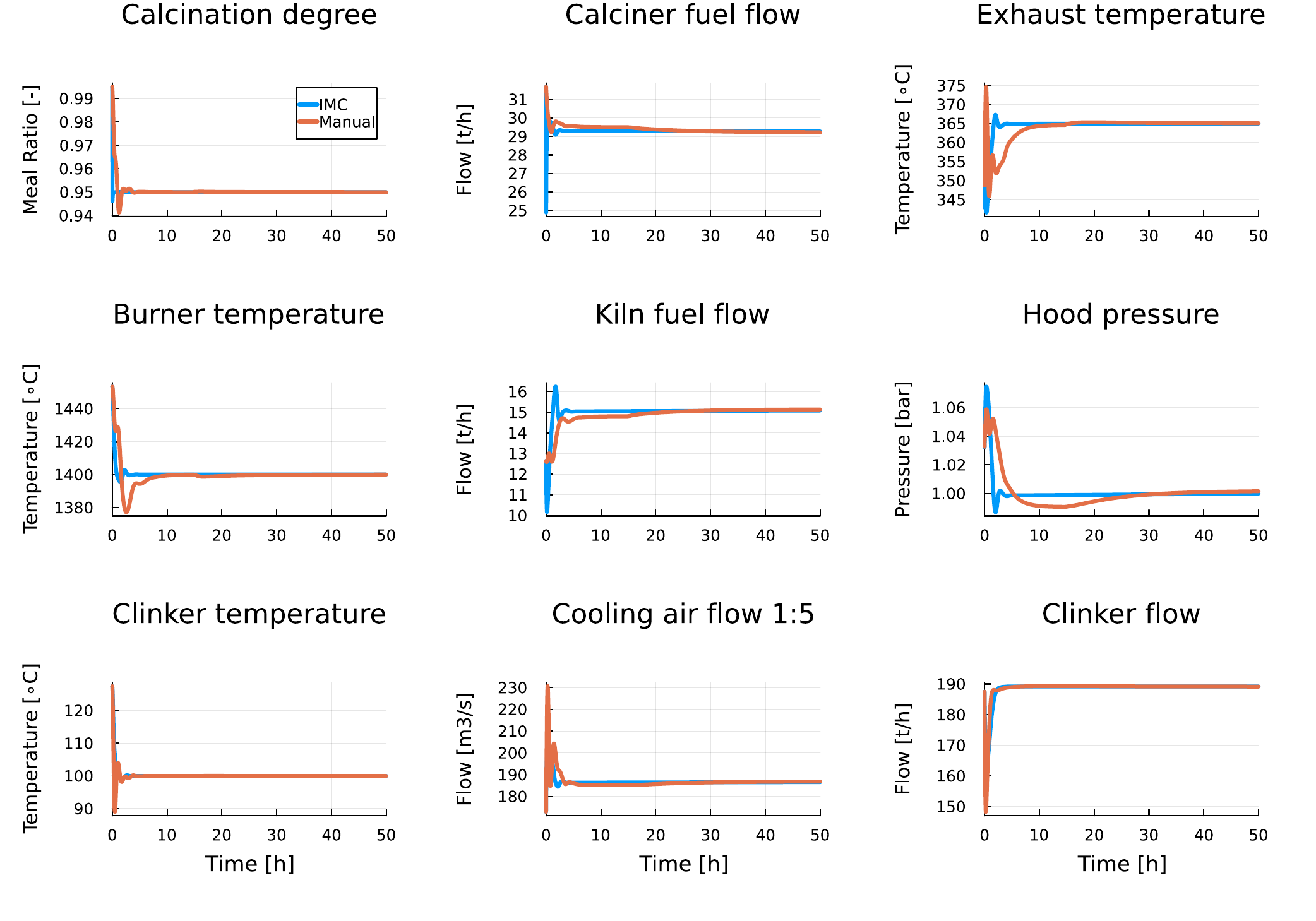}
    
    \vspace{0.3cm}
    
    \includegraphics[width=1\linewidth]{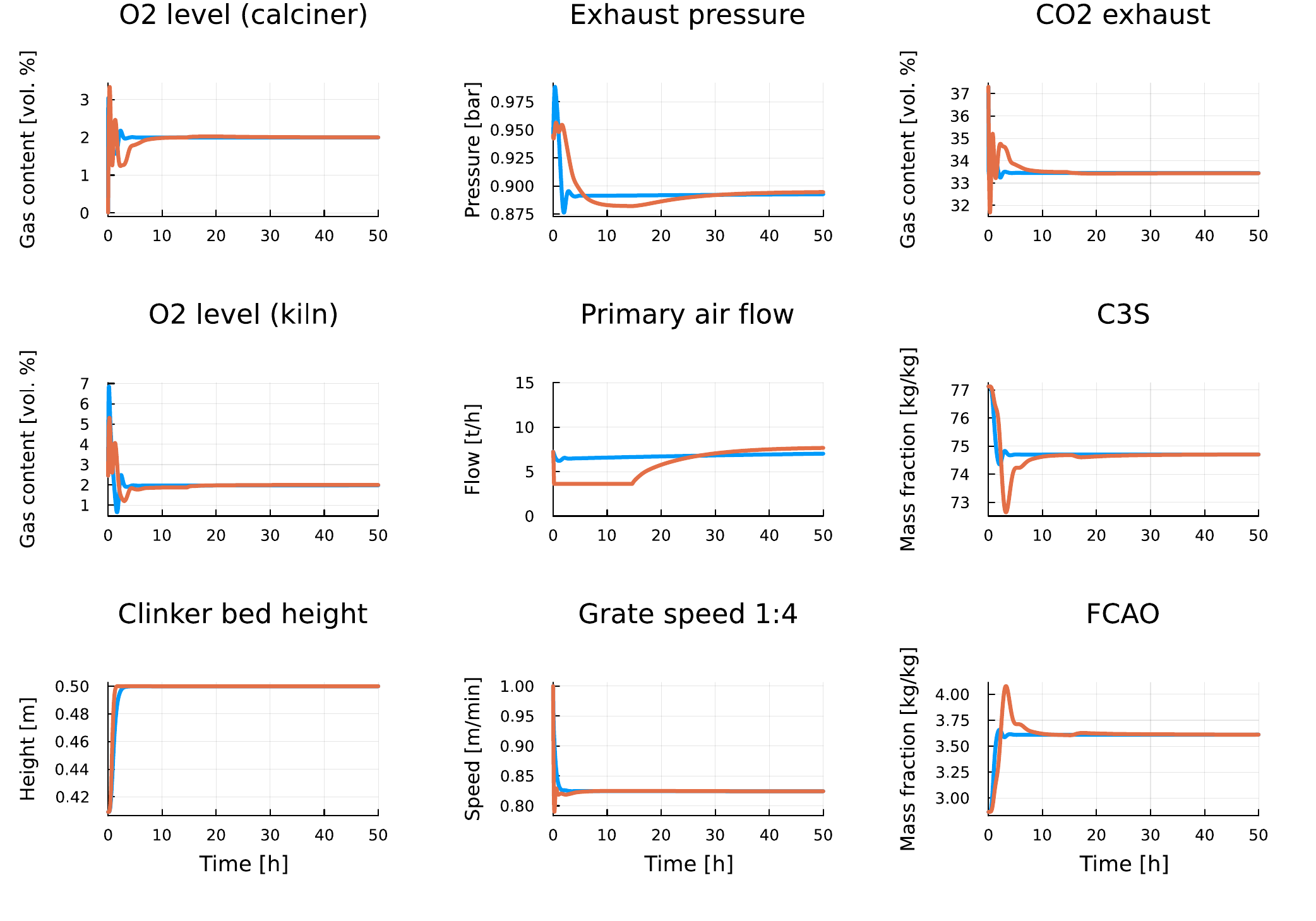}
    \caption{50-hour closed-loop simulations using the IMC-PIs and the manually tuned PIs.
    The first column shows the CVs, the second column shows the MVs, while the third column shows KPIs of interest for the process, such as clinker content (C3S, FCAO [free lime]).}
    \label{fig:IMC_tuned_long_1}
\end{figure}

\begin{figure}
    \centering
    \includegraphics[width=1\linewidth,trim={0cm 2cm 0cm 0.0cm},clip]{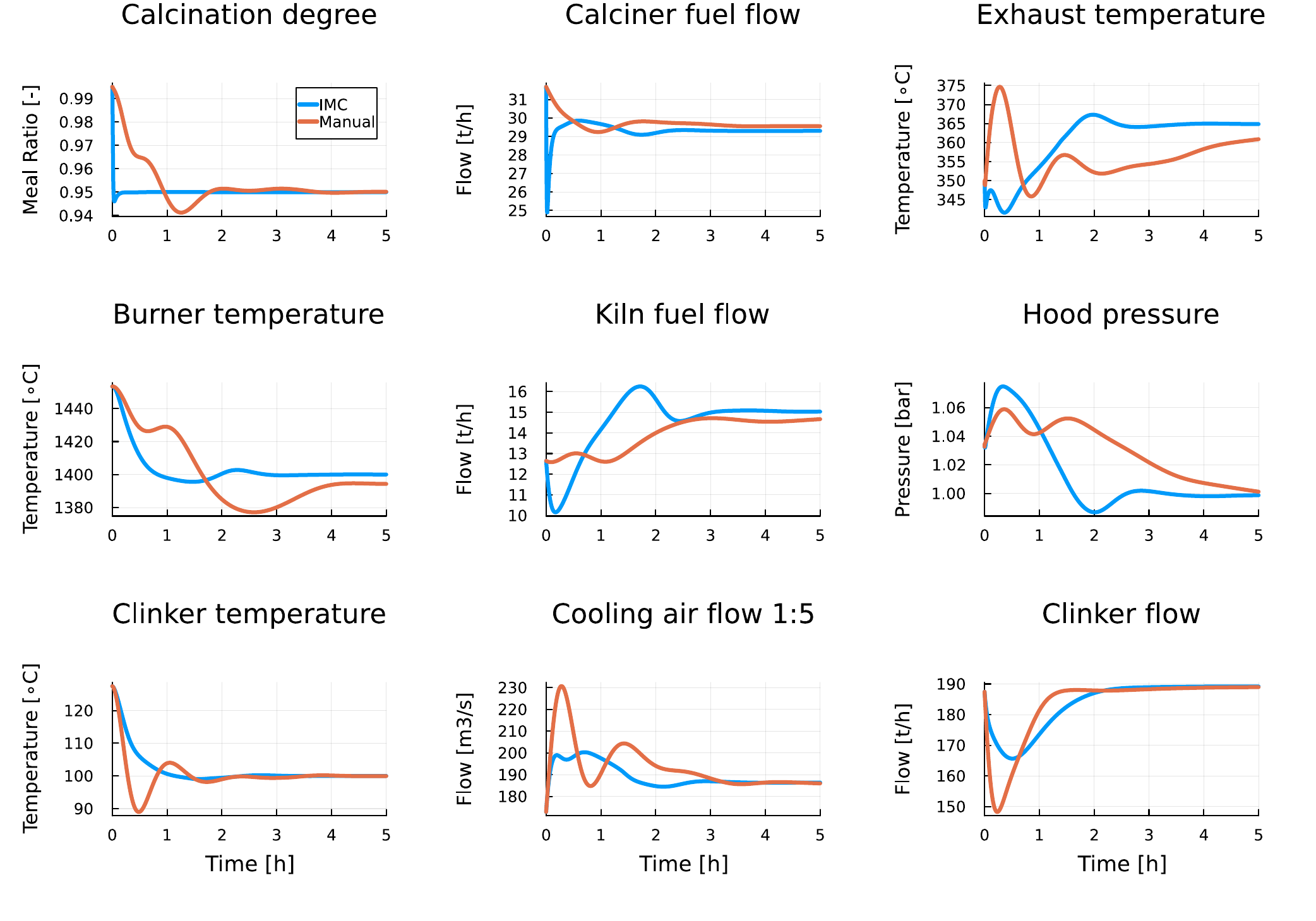}
    
    \vspace{0.3cm}

    \includegraphics[width=1\linewidth]{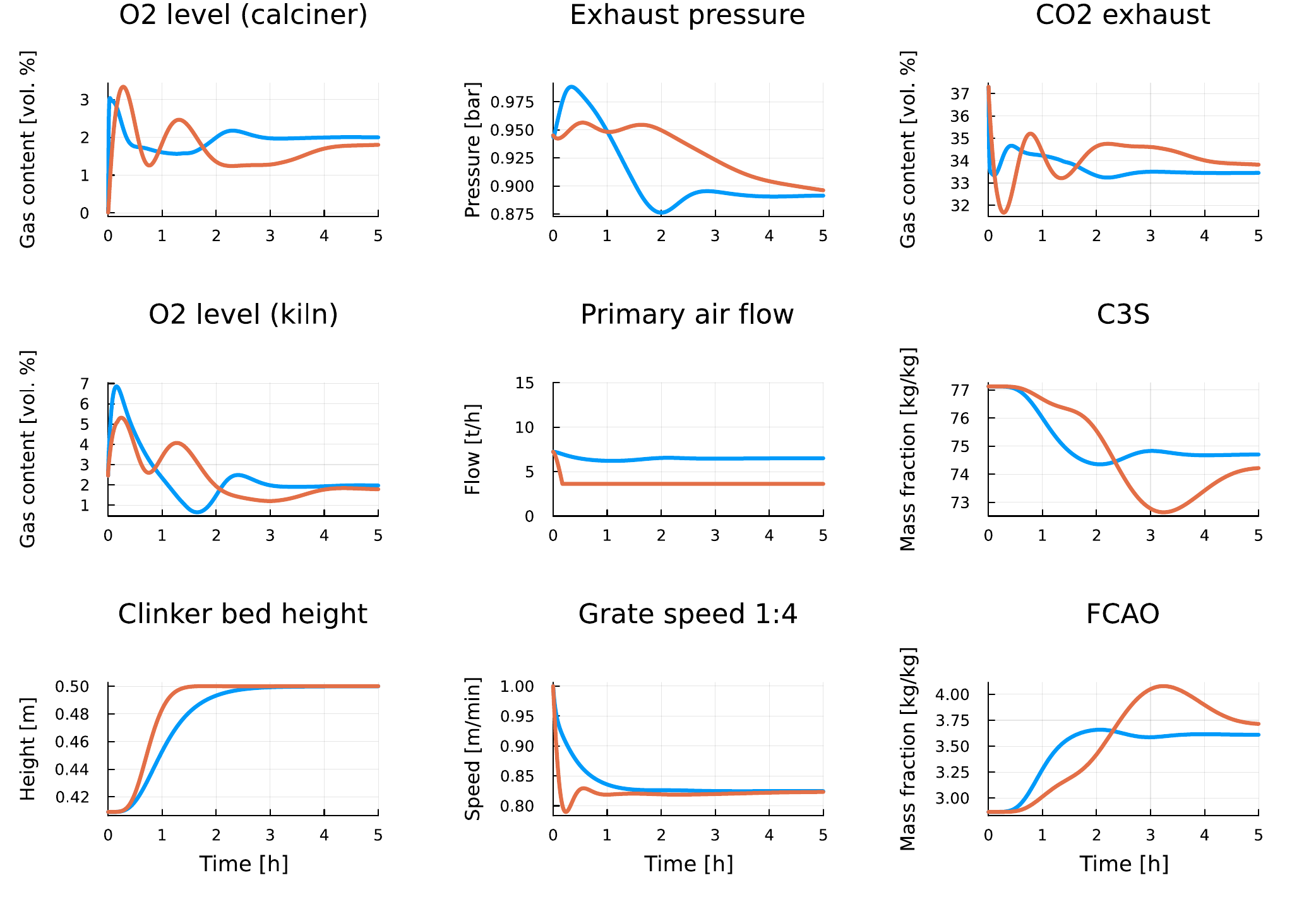}    
    \caption{The first 5 hours of the closed-loop simulations in Figure \ref{fig:IMC_tuned_long_1}. Clarifying the faster settling of the IMC-PI.
    The columns show the CVs, MVs, and selected KPIs.}
    \label{fig:IMC_tuned_short_1}
\end{figure}

  
Figure~\ref{fig:IMC_tuned_long_1} shows the CVs, the MVs, and selected process KPI values for simulations with the PI controllers with manaul tuning and IMC-based tuning. Figure~\ref{fig:IMC_tuned_short_1} is the same simulations during the first 5 hours. Figures~\ref{fig:IMC_tuned_long_1} and \ref{fig:IMC_tuned_short_1} show that the IMC tuned decentralized PI-controller provides an overall smoother and faster operation than the manually tuned decentralized PI-controller. The burner temperature and calcination degree settle fast for the IMC-PI, while the manually tuned PI takes several hours to settle. The responses are also more similar to first/second-order responses than for the manually tuned PI.
We also see that given the primary air's less integral action, it does not reach saturation, allowing the other PIs to settle without a sudden change when the saturation finishes.

This gives a good demonstration of how access to a simulator model can improve control design and tuning for the cement process.

\section{RGA}\label{sec:RGA}
In this section, we utilize the model to evaluate the control structure.
The pairing of MVs and CVs in the decentralized PI controller can be evaluated using the relative gain array (RGA) method described by \cite{Jorgensen:Jorgensen:2000cace} and \cite{skogestad2005multivariable}.

The RGA method computes a matrix, $\Lambda$ from a transfer function matrix $G$ at a given frequency $\omega$,
\begin{equation}
    \Lambda_{\omega}(G) = G(i \omega)\circ(G(i \omega)^{-1})^T.\footnote{$\circ$: the schur product or element-wise multiplication}
\end{equation}
When $[\Lambda_{\omega}(G)]_{i,j} = 1$, the pairing is ideal. The pairing is usually evaluated using the relative interaction (RIA),
\begin{equation}
    [\Phi_\omega]_{i,j} = \frac{1}{[\Lambda_{\omega}(G)]_{i,j}}  -1.
\end{equation}
In the analysis, we apply a sequential evaluation of the pairings. At each step, we find the pairing with the minimum absolute relative interaction,
\begin{equation}
    (i,j) = \arg \min(|\Phi_0|).
\end{equation}
The corresponding row and column of the pair are removed from $\Phi_0$, and we repeat to find the next pair. Alternatively, one can minimize the sum of relative interactions by solving an assignment optimization problem \citep{Jorgensen:Jorgensen:2000cace}.

Table~\ref{tab:RGA_0} shows the RGA matrix, $\Lambda_0$, evaluated at the steady-state frequency, $\omega = 0$. The clinker outflow, $F_{clink}$, and the raw meal feed inflow, $F_{feed}$, are appended to the CVs and MVs, respectively. Table~\ref{tab:RGA_0} also shows the (MV-CV)-pairing for the simulations and the suggested pairing with the RGA/RIA method. In both cases, the grate belt speed and feed flow have the same pairs with almost ideal RGA value.  
As the remaining PI-loop pairs have high values, they are less ideal pairings than those provided by the RGA method. 
Though relatively, the $T_{clinker}$-$F_{f,Ca}$ pairing is worse than the $T_{clinker}$-$F_{cool}$ pair is.
\begin{table}[]
    \centering
     \scriptsize
    \caption{RGA matrix. PI-loop pairs (blue,orange). RGA-based pairs (red,orange).}
    \begin{tabular}{|c|ccccccc|} \hline
          \text{CV\textbackslash MV}     & $v_{grate}$ &$P_{ph}$& $F_{feed}$ &  $F_{f,Ca}$   &$F_{f,K}$ & $F_{1st}$& $F_{cool}$\\ \hline
        $X_{O2,Ca}$      &-7e-6        & \cellcolor{cyan!30}5.20   &  -0.08     & -1.13         &     \cellcolor{red!30}1.61 & -2.17    &   -2.43\\
        $X_{CaO}$ & 1e-5        & -6.62  &  1.30      &  \cellcolor{cyan!30}13.4         &   -14.4  &  \cellcolor{red!30}3.87    &  3.51\\
        $X_{O2,K}$       & 8e-5        & -3.56  & -0.02      &  -0.01        &  -4.23   &    \cellcolor{cyan!30}7.14  &   \cellcolor{red!30}1.69\\
       $T_{s,burn}$      &-0.01        &  \cellcolor{red!30}8.54  & -1.38      & -11.7         &  \cellcolor{cyan!30}19.4    &   -8.35  &  -5.51\\
       $F_{clink}$       & 1e-3        & -0.17  &  \cellcolor{orange!30}1.04      &   2e-3        &   0.01   &   0.04   &   0.08\\
       $T_{clinker}$     &-2e-6        & -2.41  &  0.18      &   \cellcolor{red!30}0.42        &  -1.26   &  0.45    &  \cellcolor{cyan!30}3.61\\
       $h_{bed}$         & \cellcolor{orange!30}1.01        &  0.02  & -0.04      &  0.02         &  -0.08   &  0.02    &   0.05 \\ \hline
    \end{tabular}
    \label{tab:RGA_0}
\end{table}

If the uniform cooling air flow assumption is changed to 3 cooling air flow rates instead, the pairing changes to ($F_{feed}$-$F_{clink}$, 0.90), ($F_{cool,1}$-$X_{O2,Ca}$, 1.05), ($P_{ph}$-$X_{CaO}$, 1.11), ($F_{cool,2:3}$-$T_{s,burn}$, 1.58), ($F_{1st}$-$X_{O2,K}$,  4.91), ($F_{cool,4:5}$-$T_{clinker}$, 1.05), and 
 ($v_{grate}$-$h_{bed}$, 1.01). The extra inputs provide more ideal pairings, though the fuel flows are not in any pairs. The $T_{clinker}$-$F_{cool,4:5}$ pair is similar to the original cooling-flow pair used in the PI-loops, supporting the rationale for the initial choices.
 
\begin{figure}
    \centering
    \includegraphics[width=\linewidth,trim={0cm 0.8cm 0cm 0.1cm},clip]{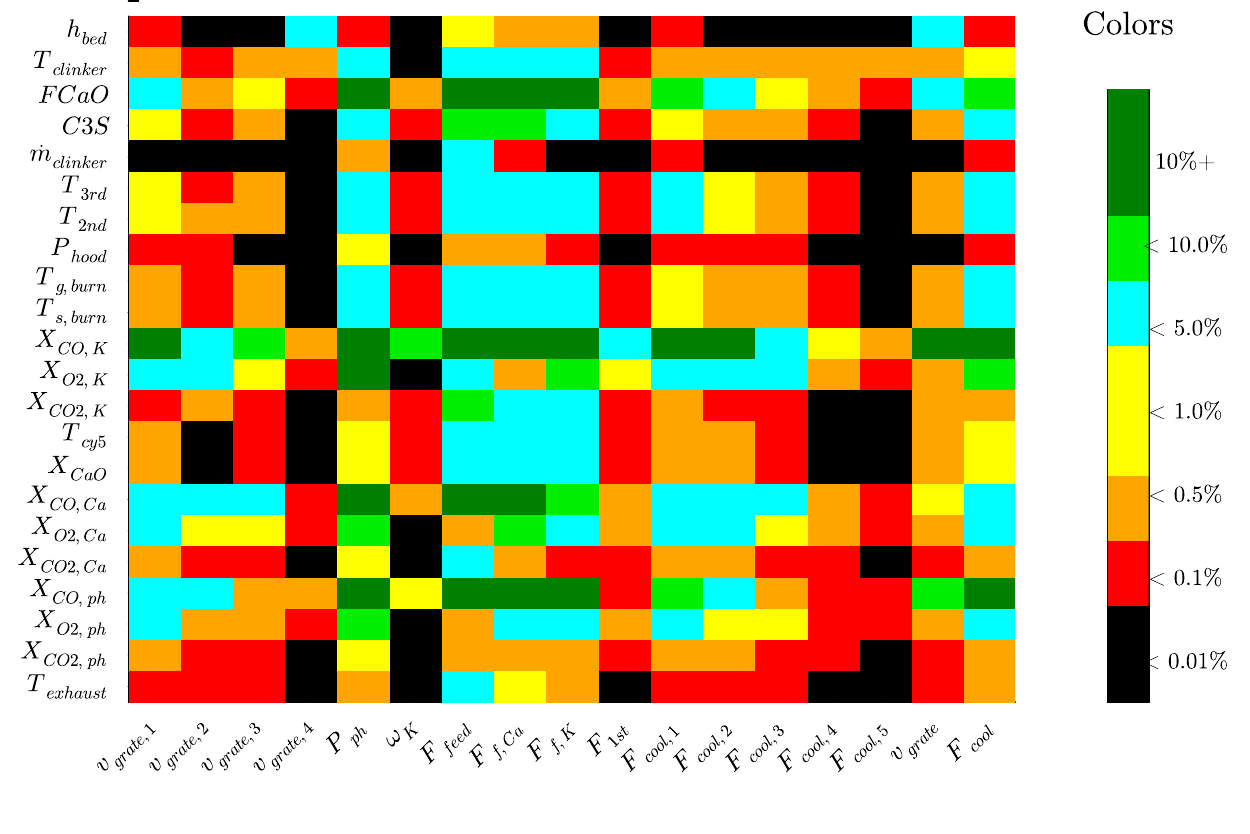}
    \caption{Responsiveness of CVs to MVs. Maximum relative change of CVs to steps in the MVs. Colored by discrete percentage ranges.}
    \label{fig:stepchange}
\end{figure}

Using the step response tests, we can illustrate the effective changes in each CV of the system against each MV.
Figure~\ref{fig:stepchange} shows the categorized percentage change of each step response.
In the RGA above, $T_{s,burn}$ was paired with different MVs depending on the MVs available. Figure~\ref{fig:stepchange} shows that $T_{s,burn}$ is most impacted by $P_{ph}$, $F_{meal}$, $F_{f,Ca}$, $F_{f,K}$, and the cooling air. This fits with the initial pairing, $F_{f,K}$, and the first RGA pairing, $P_{ph}$. The second RGA pairing chooses a less changing pairing, thus a less sensitive interaction.

\section{Conclusion}\label{sec:Conclusion}
We used a simulation model for the pyro section in a cement plant to design and tune a decentralized PI controller.
We demonstrated the use of the model to obtain linear models that are used for RGA-based control structure selection and IMC-based tuning. The linear models describe the process well in a small neighborhood around the steady state operating point. Such models are difficult to obtain in practical cement plant operation, but they can readily be obtained from a high-fidelity simulator. The paper is one step towards model-based design of controllers for cement plants. 
Future work considers the use of the identified linear model for model predictive control (MPC) design.





\bibliography{ref/biblio_paper}             
                                                   







\end{document}